\documentclass[conference]{IEEEtran}
\IEEEoverridecommandlockouts
\usepackage{cite}
\usepackage{amsmath,amssymb,amsfonts}
\usepackage{algorithmic}
\usepackage{graphicx}
\usepackage{textcomp}
\usepackage{xcolor}
\usepackage{xfrac}
\usepackage{hyperref}
\usepackage[utf8]{inputenc} 
\usepackage[T1]{fontenc}
\usepackage{url}
\usepackage{ifthen}
\usepackage{cite}
\usepackage{graphicx}
\usepackage{amsfonts}
\usepackage{amssymb}
\usepackage{bbm}
\usepackage{subfigure}
\usepackage{diagbox}
\usepackage{tikz}
\usepackage{booktabs,multirow}
\usepackage{array}
\usepackage{caption}
\usetikzlibrary{decorations.markings}

\tikzset{->-/.style={decoration={
  markings,
  mark=at position #1 with {\pgftransformscale{1.5}\arrow{>}}},postaction={decorate}}}
 
 \tikzset{-<-/.style={decoration={
  markings,
  mark=at position #1 with {\pgftransformscale{1.5}\arrow{<}}},postaction={decorate}}}

\usepackage[ruled,vlined]{algorithm2e}
\def\BibTeX{{\rm B\kern-.05em{\sc i\kern-.025em b}\kern-.08em
    T\kern-.1667em\lower.7ex\hbox{E}\kern-.125emX}}

\newcommand{\figref}[1]{Fig.~\ref{#1}}
\newcommand{\tabref}[1]{Table~\ref{#1}}

\begin{document}

\title{Interpreting Neural Min-Sum Decoders \\
\thanks{To appear in proceedings of IEEE International Conference on Communications (ICC), 2023 }
}

\author{\IEEEauthorblockN{Sravan Kumar Ankireddy, Hyeji Kim}
\IEEEauthorblockA{Department of Electrical and Computer Engineering \\
\text{University of Texas at Austin}\\
}
\vspace{-6mm}
}
\maketitle
\begin{abstract} In decoding linear block codes, it was shown that noticeable reliability gains can be achieved by introducing learnable parameters to the Belief Propagation (BP) decoder. Despite the success of these methods, there are two key open problems. The first is the lack of interpretation of the learned weights, and the other is the lack of analysis for non-AWGN channels. In this work, we aim to bridge this gap by providing insights into the weights learned and their connection to the structure of the underlying code. We show that the weights are heavily influenced by the distribution of short cycles in the code. We next look at the performance of these decoders in non-AWGN channels, both synthetic and over-the-air channels, and study the complexity vs. performance trade-offs, demonstrating that increasing the number of parameters helps significantly in complex channels. Finally, we show that the decoders with learned weights achieve higher reliability than those with weights optimized analytically under the Gaussian approximation. 
\end{abstract}

\section{Introduction}\label{sec:intro}

Short length block codes are an attractive choice for use cases with stringent latency requirements \cite{ccsds}. However, they suffer from poor error correction performance. One key reason for this is the presence of short cycles in the Tanner graph of the code, making the iterative decoding sub-optimal. Unfortunately at such short lengths, codes designed without cycles have worse error correction performance\cite{ryan2009channel}. Additionally, when fading channels are considered, the Log Likelihood Ratio (LLR) computation suffers from errors because of imperfect Channel State Information (CSI) and equalization, which makes the Belief Propagation (BP) decoder sub-optimal \cite{yazdani2011}. Hence modifications to the traditional BP decoder are necessary to improve its performance. 

In \cite{nachmani2016}, the authors demonstrate the advantage of using neural networks for improving the BP decoder by placing multiplicative weights along the edges of the Tanner graph structure and unrolling the iterations, resulting in a neural BP decoder.  
In \cite{noms2017}, the authors show that instead of using multiplicative weights, comparable gains can be achieved by using offset factors combined with min-sum approximation, which reduces the implementation cost. Later in \cite{nachmani2018}, the authors explore the possibility of reducing the number of weights needed by entangling them across iterations in a recurrent fashion and show that the performance is still comparable to the fully parameterized model. More recently, the authors in \cite{lian2019} propose entangling the weights not only across iterations but also across the edges. To compensate for the performance loss, a shallow neural network referred to as Parameter Adapter Network (PAN), is used to select different weights for different SNRs regions resulting in a performance very close to \cite{nachmani2018}. Recently in \cite{buchberger2020pruning}, the authors propose using the weights from neural BP decoders to prune an over-complete parity check matrix to find the most important check nodes. Authors in \cite{nachmani2022} explore sparse constraints on node activations and use knowledge distillation to improve the neural decoders. 

Despite the success of these works, there are two key open problems that we answer in this work. The first is the interpretation of the learned weights. While it has been conjectured that the use of normalization or offsets to modify the beliefs helps mitigate the detrimental effects of short cycles \cite{nachmani2016,noms2017}, there is no supporting evidence. We present a first empirical evidence that the weights learned are directly related to the short cycles present in the code and investigate how they help in improving the reliability of the posterior LLRs.

The second is the analysis and trade-offs associated with the complexity of neural min-sum decoders for channels beyond the Additive White Gaussian Noise (AWGN) channels. Existing work has focused on AWGN channels
~\cite{nachmani2016,noms2017,nachmani2018,lian2019, wang2021}.  
We demonstrate that the trade-offs are quite different on complicated channels (including the over-the-air channel) from AWGN channels, showing a need for an adaptive selection of model complexity based on channel conditions. Our main contributions are as follows: 
\begin{itemize}
    \item \emph{Interpretation:} We interpret the gains provided by neural min-sum decoders in a two-phase fashion, first by correcting for {\em cycles} in the code structure and then providing additional gains by correcting for {\em channel effects} (Section~\ref{sec:int}).
 
     \item \emph{Complexity Trade-off:} We explore the complexity vs. performance trade-off for neural min-sum decoders across various channels including both synthetic and {\em over-the-air} channels. 
     We show that complex channels and over-the-air channels require more weights to fully realize the possible gains, while fewer weights suffice for simple channels. To the best of our knowledge, we are the first to evaluate and study neural min-sum decoders in non-AWGN channels
      (Section~\ref{sec:ent}).  

    \item \emph{Robustness and Adaptivity:} We show that the learned weights are fairly robust to channel variations \textit{i.e,} the decoder trained on one channel still outperforms the classical decoders on other channels. Furthermore, any loss in performance due to the change in channel conditions can be easily recovered with small fine-tuning (Section~\ref{sec:robustness}).

    \item \emph{Gaussian Approximation:} 
    Finally, we propose an analytical approach for finding the weights using Gaussian approximation and compare the neural min-sum decoders and analytical decoders in terms of reliability. We show that for complicated channels, neural decoders lead to much better performance than the analytically driven weights under the Gaussian approximation. 
    (Section~\ref{sec:GA}).
    
\end{itemize}

\section{System Model}\label{sec:sysmodel}
In this work, we consider linear block codes. A $(N,K)$ block code maps a message of length $K$  to a codeword of length $N$ and is uniquely described by its parity check matrix $\mathbb{H}$ of dimensions $(N-K) \times N$, where the rate of the code is $R=\sfrac{K}{N}$. The linear block code can be also represented using a bipartite graph, known as the Tanner graph, which can be constructed using its parity check matrix $\mathbb{H}$. The Tanner graph consists of two types of nodes. We refer to them as Check Nodes (CN) that represent the parity check equations and Variable Nodes (VN) that represent the symbols in the codeword. There is an edge present between a check node $c$ and variable node $v$ if $\mathbb{H}(c,v) = 1$.

We consider a system with Binary Phase Shift Keying (BPSK) modulation that transmits a random vector $X \in \{-1,1\}^N$, where $N$ is the code length. The modulated signal is then passed through a channel to receive $Y$ as
\begin{equation*}
    Y = \text{Channel}(X) + W,
\end{equation*}
where Channel($X$) applies the effect of channel on the transmit vector $X$ and $W$ is the noise at the receiver. As a special case, when the channel is AWGN the LLR at the receiver for $v^{\text{th}}$ symbol can be calculated as
\begin{equation*}
    l_v = \frac{p(Y_v = y_v | X_v = 1)}{p(Y_v = y_v | X_v = -1)} = -\frac{2y_v}{\sigma^2},
\end{equation*}
where $\mbox{Var}(W) = \sigma^2$ is the noise variance.

Apart from the AWGN channel, we also consider multi-path fading propagation channels from the 3GPP specifications \cite{3gppts36101}. Specifically, we consider a multi-path fading ETU channel, which has a high delay spread. We use the MATLAB LTE Toolbox to transmit and receive data. The multi-path delay profiles can be found in~\tabref{ETUTable}. Finally, we also test the decoder over-the-air by transmitting and receiving using a USRP N200 SDR setup in a non-line-of-sight (NLOS) multi-path environment. More comprehensive results and the source code can be found at \href{https://github.com/sravan-ankireddy/nams}{https://github.com/sravan-ankireddy/nams}.

\section{Background}\label{sec:background}
In this section, we review the {\em min-sum} variant of the BP decoder and neural decoders with augmented weights. \vspace{.1em}

\textbf{Min-sum decoding.\ } 
The BP decoder is an iterative soft-in soft-out decoder that operates on the Tanner graph to compute the posterior LLRs of the received vector, also referred to as beliefs. In each iteration, the check nodes and the variable nodes process the information to update the beliefs passed along the edge. Operating in such an iterative fashion allows for incremental improvement in the estimated posteriors.

During the first half of iteration $t$, at the VN $v$, the received channel LLR $l_v$ is combined with the remaining beliefs $\mu^{t-1}_{c',v}$ from check node to calculate a new updated belief, to be passed to the check nodes in next iteration. Hence, the message from VN $v$ to CN $c$ at iteration $t$ can be computed as
\begin{equation}
    \mu^t_{v,c} = l_v + \sum_{c' \in N(v) \setminus c} \mu^{t-1}_{c',v} \mbox{          }, 
\end{equation}
where $N(v)\setminus c$ is the set of all check nodes connected to variable node $v$ except $c$, as illustrated below. 
\begin{center}

\begin{tikzpicture}[
  vnode/.style = {shape=circle,draw,minimum size=1em,fill=blue!100},
  cnode/.style = {shape=rectangle,draw,minimum size=1em,fill=red!100},
  edge/.style = {-},
  ]
  \node[] (ch) at (0,-1.5) {};
  \node[vnode,label=right:$v$] (c) at (0,0) {};
  \node[cnode,label=right:$c_i$] (ci) at (-3,2) {};
  \node[cnode,label=right:$c_j$] (cj) at (-1,2) {};
  \node[cnode,label=right:$c_k$] (ck) at (1,2) {};
  \node[cnode,label=right:$c_l$] (cl) at (3,2) {};

  \draw[edge, -<-=0.5] (c) to node[right, xshift=1mm] {$l_v$} (ch);
  \draw[edge, red, thick, ->-=0.5] (c) to node[left, xshift=-1mm, yshift=-1mm] {$\mu^t_{v,c_i}$} (ci);
  \draw[edge, -<-=0.5] (c) to node[right, xshift=-1mm, yshift=3mm] {$\mu^{t-1}_{c_j,v}$} (cj);
  \draw[edge, -<-=0.5] (c) to  node[right, xshift=1mm, yshift=3mm] {$\mu^{t-1}_{c_k,v}$} (ck);
  \draw[edge, -<-=0.5] (c) to node[right, xshift=5mm, yshift=3mm] {$\mu^{t-1}_{c_l,v}$} (cl);
\end{tikzpicture}
\end{center}

During the latter half of the iteration $t$, at the CN $c$, the message from the CN to any VN is calculated based on the criterion that the incoming beliefs  $\mu^t_{v',c}$ at any check node should always satisfy the parity constraint. To reduce the computational complexity of BP decoding, a hardware friendly variant known as \textit{min-sum} approximation is used in practice. The message from CN $c$ to VN $v$ at iteration $t$ is given by
\begin{equation}\label{eq:cn_ms}
    \mu^t_{c,v} =  \min_{v' \in M(c) \setminus v}  ( |\mu^t_{v',c}|) \prod_{v' \in M(c)  \setminus v} \text{sign} \left(\mu^t_{v',c} \right).
\end{equation}
where $M(c) \setminus v$ is the set of all variable nodes connected to check node $c$ except $v$ , as illustrated below. 

\begin{center}
\begin{tikzpicture}[
  vnode/.style = {shape=circle,draw,minimum size=1em,fill=blue!100},
  cnode/.style = {shape=rectangle,draw,minimum size=1em,fill=red!100},
  edge/.style = {-},
  ]
  \node[cnode,label=right:$c$] (c) at (0,0) {};
  \node[vnode,label=right:$v_i$] (vi) at (-3,-2) {};
  \node[vnode,label=right:$v_j$] (vj) at (-1,-2) {};
  \node[vnode,label=right:$v_k$] (vk) at (1,-2) {};
  \node[vnode,label=right:$v_l$] (vl) at (3,-2) {};

  \draw[edge, blue, thick, ->-=0.5] (c) to node[left, xshift=-1mm, yshift=1mm] {$\mu^t_{c,v_i}$} (vi);
  \draw[edge, -<-=0.5] (c) to node[right, xshift=-1mm, yshift=-2mm] {$\mu^{t}_{v_j,c}$} (vj);
  \draw[edge, -<-=0.5] (c) to  node[right, xshift=1mm, yshift=-2mm] {$\mu^{t}_{v_k,c}$} (vk);
  \draw[edge, -<-=0.5] (c) to node[right, xshift=5mm, yshift=-2mm] {$\mu^{t}_{v_k,c}$} (vl);
\end{tikzpicture}
\end{center}

Finally, at the end of iteration $t$, we combine all the incoming beliefs to estimate the posterior belief as
\begin{equation}
    o^{t}_{v} = l_v + \sum_{c' \in N(v)} \mu^{t}_{c',v} \mbox{        }.
\end{equation}

\textbf{Neural min-sum decoding.\ }  While the min-sum approximation simplifies the computation, it also comes with a loss in performance. It is readily shown in \cite{chen2002_nms} that the min-sum approximation is always greater than the true LLR from BP. The neural min-sum algorithm improves the performance of the min-sum decoder by introducing trainable \textit{weights} along the edges of the Tanner graph\cite{noms2017,nachmani2018}. 

Depending on the choice of correction, \eqref{eq:cn_ms} can be modified to produce Neural Normalized Min-Sum (NNMS) and Neural Offset Min-Sum (NOMS) algorithms, given by
\begin{equation}\label{eq:nnms}
    \mu^t_{c,v} =  \alpha_{v^\star,c}^t |\mu^t_{v^\star,c}| \prod_{v' \in M(c)  \setminus v} \text{sign} \left(\mu^t_{v',c} \right),
\end{equation}
\begin{equation}\label{eq:noms}
\begin{split}
    \mu^t_{c,v} =   \mbox{max} \left( |\mu^t_{v^\star,c}| - \beta_{v^\star,c}^t, 0 \right)  \prod_{v' \in M(c)  \setminus v} \text{sign} \left(\mu^t_{v',c} \right),
\end{split}
\end{equation}
respectively, where $v^\star = \underset{v' \in M(c) \setminus v}{\arg\min} |\mu^t_{v',c}|$. The coefficients $\alpha_{v^\star,c}^t$ and $\beta_{v^\star,c}^t$ denote trainable normalization and offset factors respectively, corresponding to the edge connecting variable node $v^\star$ to check node $c$ in iteration $t$. These weights are then learned via stochastic gradient descent algorithms using off-the-shelf deep learning frameworks to provide noticeable reliability improvements.


\section{Interpreting the Neural Min-Sum decoders }\label{sec:int}
While neural decoders achieve  impressive gains, it is natural to wonder about the reason for these gains. In \cite{nachmani2016,nachmani2018,lian2019}, it has been conjectured that the weights mitigate the effect of cycles and thus improve the performance but evidence for the same is lacking. We propose that the neural min-sum decoders provide gains in two phases. The first is, as conjectured previously, gains are achieved by correcting for the false overestimates introduced by the cycles. In this regard, we provide the first empirical evidence that the weights are closely associated with the cycles in the graph. Additionally, by extending the analysis to non-AWGN channels, we show that neural networks also learn to correct the channel effects and provide additional gains.

For ease of analysis, we consider the NNMS version of the neural min-sum decoder. \textcolor{black} {We consider BCH(63,36) code for our primary analysis, consistent with previous works~\cite{noms2017,nachmani2016,lian2019}. The performance of these codes is studied well in literature~\cite{channelcodes} and also similar codes have been used in practice for space data systems~\cite{ccsds}. While the BCH family of codes makes it easier to realize the gains from modifications to the min-sum decoder because of the inherent sub-optimality of using the min-sum decoder for BCH codes, the same principles and analysis can be applied to any linear block codes.} Additionally, we follow the same architecture and training methodology as \cite{noms2017}. \vspace{.5em}

\begin{figure}[!ht]
    \vspace{-2mm}
    \centering
 	\includegraphics[width=1.0\linewidth]{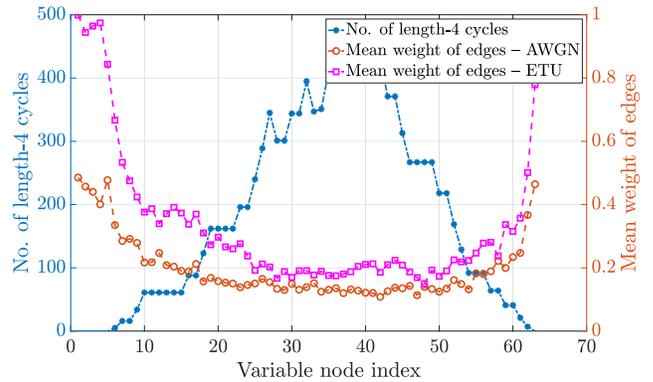}
 	\captionsetup{font=small}
 	\caption{Weights vs Cycles: The mean weights, across the edges and iterations, learned by NNMS at each check node (for BCH(63,36)) are inversely proportional to the number of short cycles present. Additionally, we observe a noticeable difference in weights between AWGN and ETU channels.}
 	\label{fig:cycVsWt}
\end{figure}

\textbf{Connection to cycles.\ } In order to study the connection between cycles in the graph and the weights learned, we enumerate the number of length-4 cycles present at each variable node and compare them against the mean of weights across all corresponding edges and iterations.  We see from \figref{fig:cycVsWt} that the weights are inversely proportional to the number of cycles present at each variable node for both AWGN and ETU channels, demonstrating that the learned weights impose higher correction in the presence of a higher number of cycles.

To investigate this further, we look at three different rates of length 63 BCH codes, which have a different number of length-4 cycles. For each code, we again measure the mean weight across all edges and iterations.  From \tabref{cyclesAndWeights}, we can see that the trend holds true, with BCH(63,30) having the highest number of cycles and hence the least mean weight, while the default weight in traditional min-sum decoder would be 1. We validate this for both AWGN and ETU channels, thus showing strong evidence that the weights being learned are indeed commensurate with the cycles present.

\begin{table}[!htp]\centering
\vspace{2mm}
\captionsetup{font=small}
\scriptsize
\begin{tabular}{lrrrr}\toprule
\multirow{2}{*}{Code} &\multirow{2}{*}{Length-4 cycles} &\multicolumn{2}{c}{\text{Average weight}} \\\cmidrule{3-4}
& &AWGN\hspace{2mm} &ETU \\\midrule

BCH (63,30) &10122\hspace{4mm} &0.2632\hspace{2mm} &0.3134 \\
BCH (63,36) &5909\hspace{5mm} &0.2987\hspace{2mm} &0.3512 \\
BCH (63,57) &1800\hspace{5mm} &0.3990\hspace{2mm} &0.5712 \\

\bottomrule
\end{tabular}

\caption{Number of short cycles present vs the weights learned. Weights are inversely proportional to the number of cycles indicating that more correction is needed when the number of cycles is higher.}
\label{cyclesAndWeights}
\vspace{1mm}
\end{table}

Since it is also conjectured that these cycles lead to a correlation between the incoming beliefs at the variable nodes~\cite{yazdani2004}, we proceed to empirically estimate and compare these correlations to better understand the effect of the weights. \vspace{.2em}

\begin{figure}[ht]
    \centering
 	\includegraphics[width=1\linewidth]{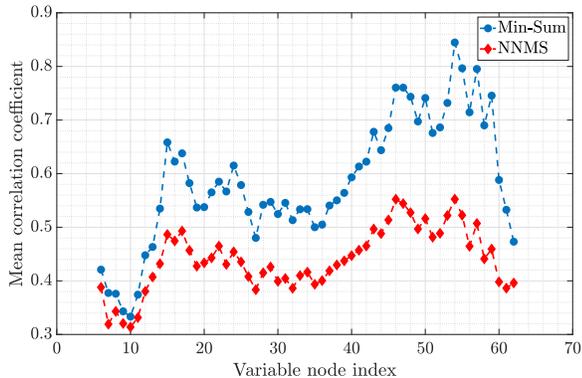}
 	\captionsetup{font=small}
 	\caption{BCH(63,36) AWGN channel at SNR 1 dB. The expected pairwise correlation coefficient between incoming messages at variable nodes is reduced across all positions.}
 	\label{fig:corr_pos}
\end{figure}

\textbf{Mitigating the correlation due to cycles.} We consider the AWGN channel for this analysis, where the correlation between incoming beliefs at nodes is because of the cycles in the Tanner graph. To find if the learned weights mitigate this effect, we measure and compare the expected pairwise  correlation coefficient between the incoming messages at the variable nodes \textit{i.e,} $\mathbb{E}[\rho_{\mu_{v,c_i}, \mu_{v,c_j}}]$, where $c_i, c_j \in N(v)$ and $i \neq j$ and the expectation is over the message and the channel. We observe from \figref{fig:corr_pos} that the correlation is reduced considerably across all the variable node positions, which supports our conjecture that the weights are improving the reliability of the posterior probabilities by reducing the correlation. 
\vspace{.2em}

Revisiting \figref{fig:cycVsWt}, even though the trends of weights across variables nodes are the same for both AWGN and ETU channels, the difference in weights between the two channels varies considerably. This indicates that the choice of channels is heavily influencing the weights learned by the decoder. To investigate this further, we study the performance of neural min-sum decoders in various channels.

\section{Channels vs. Decoder Complexity}\label{sec:ent}

\begin{figure*}[!ht]
  \centerline{\subfigure[AWGN]{  \includegraphics[width=0.5\textwidth]{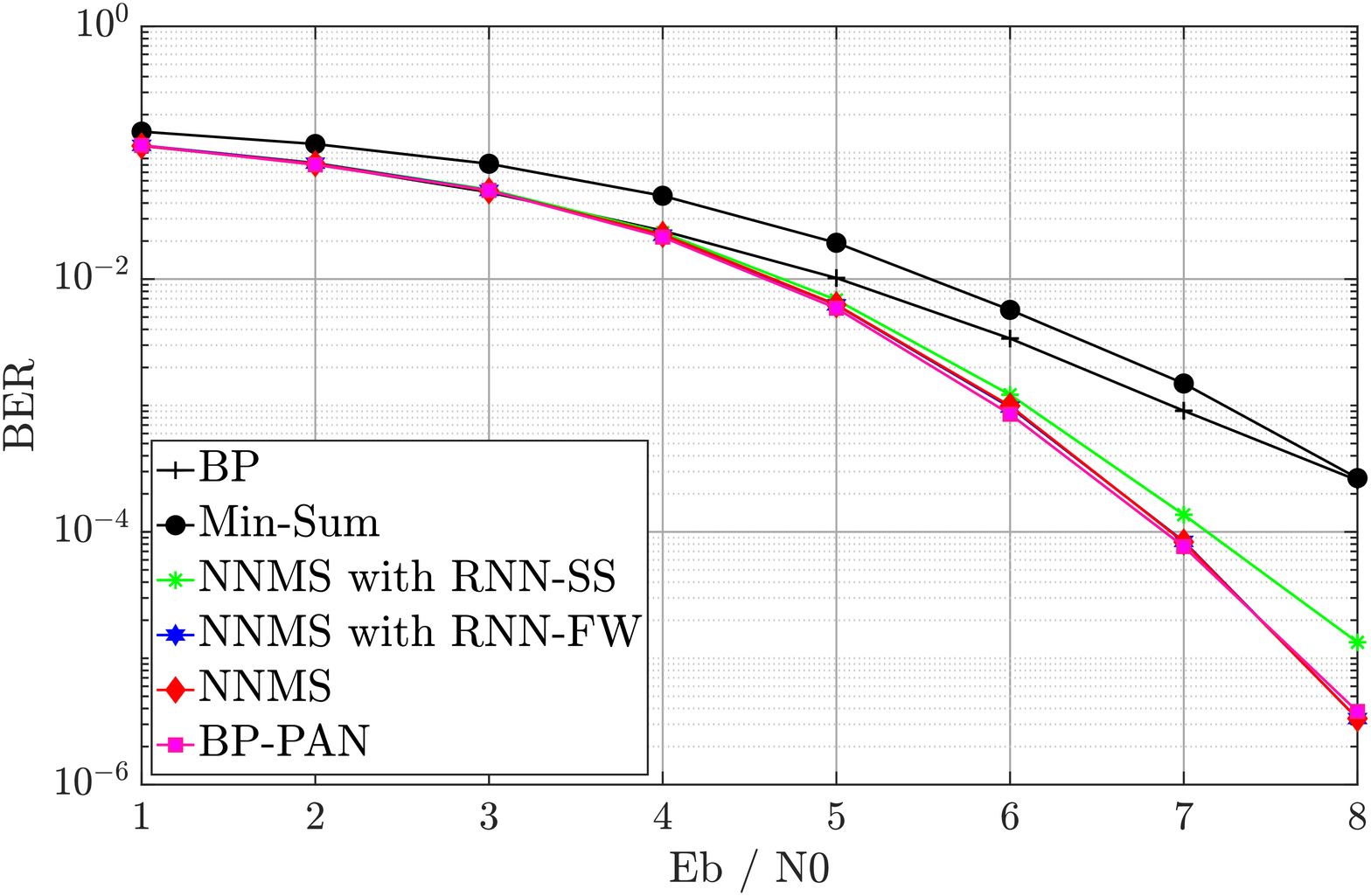}
    \label{fig:bch_63_36_AWGN}}
    \hfil
    \subfigure[ETU]{\includegraphics[width=0.5\textwidth]{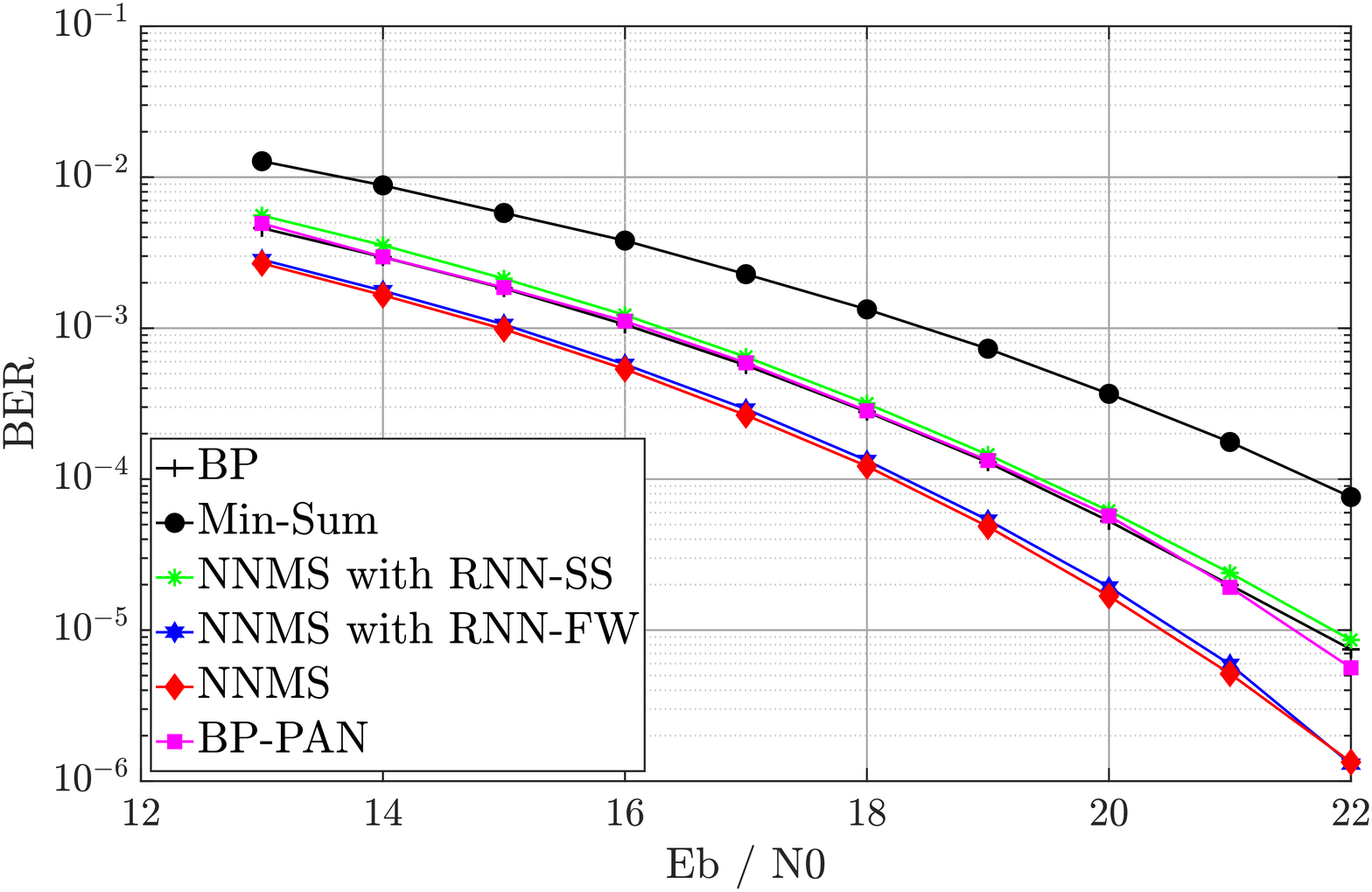}
    \label{fig:bch_63_36_ETU}}
    }
    \captionsetup{font=small}
  \caption{Effect of entangling the weights: On the left, we see that SS entanglement of weights has up to 0.3 dB of degradation at a BER of $10^{-5}$ for AWGN channel for BCH(63,36). However, for the ETU channel on the right, NNMS and RNN-FW outperform RNN-SS by up to 1.8 dB. We also see that while BP-PAN performs very well on the AWGN channel, it is unable to match with NNMS for a more complex ETU channel.}
  \label{fig:ber_ent}
\end{figure*}

While the neural min-sum decoders come with a noticeable gain in error correction performance, they also increase the memory requirement because of the large number of parameters. Hence, it is important to study the gains achieved and explore possible simplifications while maintaining little to no trade-off in performance. 

The authors in \cite{nachmani2018} propose entangling the weights across iterations to form a Recurrent Neural Network (RNN) decoder referred to as RNN-FW. Further, in \cite{lian2019}, the authors propose entangling weights not only across iterations but across edges as well, referred to as RNN-SS. This results in a loss of performance, which is compensated using a parameter adapter network that selects different weights for different SNR regions, known as BP-PAN. Finally, the authors in \cite{nachmani2018,lian2019} conclude that entangling the weights across iterations or/and edges only comes with a negligible loss. 

But, as evident from \figref{fig:cycVsWt}, the weights learned by the decoder change considerably depending on the channel. The commonly used AWGN channel might be too simple compared to more complex realistic channels, and it is not clear whether the trade-offs observed in \cite{nachmani2018,lian2019} hold true for non-AWGN channels. It is important to study the entanglement of weights not just for one channel but across various channels.

Hence we rigorously evaluate the error correction performance of different neural decoders with varying levels of entanglement across different synthetic and real channels. Based on this, we provide strong empirical evidence that having more weights is considerably beneficial in more complicated channels compared to AWGN channels.\vspace{.2em}

\textbf{AWGN channels.\ }In \figref{fig:ber_ent} we plot the Bit Error Rate (BER) performance of three variants of the NNMS decoder with varying degrees of entanglement. We choose BCH(63,36) specifically to enable direct comparison against existing neural decoders for AGWN channels~\cite{nachmani2016,noms2017,nachmani2018,lian2019}. As expected from prior work and can be seen from \figref{fig:bch_63_36_AWGN}, the performance is almost indistinguishable across the variants for the AWGN channel. \vspace{.2em}

\textbf{ETU channels.\ }In \figref{fig:bch_63_36_ETU}, we perform the same experiment for the ETU channel, where we estimate the channel at the receiver using the pilots of OFDM symbols and use MMSE equalizer from LTE MATLAB toolbox to perform equalization. We choose the ETU channel over EPA/EVA channels because of the high delay spread environment. The multi-path delay profile for the ETU channel is provided in~\tabref{ETUTable}. \vspace{.5em}

\begin{table}[!htp]\centering
\captionsetup{font=small}
\scriptsize
\begin{tabular}{cc}\toprule
\textbf{Excess tap delay (ns)} &\textbf{Relative power (dB)} \\\midrule
 0  & –1.0 \hspace{-10mm}\\
 50 & –1.0 \\
 120 &–1.0 \\
 200 &0 \\
 230 &0 \\
 500 &0 \\
 1600 &–3.0 \\
 2300 &–5.0 \\
 5000 &–7.0 \\
\bottomrule
\end{tabular}
\vspace{-0.5mm}
\caption{Multipath delay spread of ETU channel}
\label{ETUTable}
\end{table}

These multi-path components make it hard to achieve perfect equalization and the reliability of decoding is significantly impacted by the choice of entanglement, even after equalization. At a BER of $10^{-5}$, NNMS and RNN-FW outperform RNN-SS by more than 1.8 dB. \vspace{.2em}

\textbf{Bursty channels.\ }To further study the effect of channel conditions on the trade-off in performance due to entanglement, we design the following experiment. We consider a family of bursty channels in which the transmitted signal gets corrupted in two steps. First, a Gaussian noise of $\mathcal{N}(0,\sigma^2)$ is added to the signal. We then select $S$ consecutive symbols uniformly at random to add bursty noise  $\mathcal{N}(0,\sigma_b^2)$. The resultant channel can be described as 
\begin{align*}
    Y &= X + W  ,\\
    Y[j:j+S] &= Y[j:j+S] + W_b,
\end{align*}
where $W \sim \mathcal{N}(0,\sigma^2) \in \mathbb{R}^N$ and $W_b \sim \mathcal{N}(0,\sigma_b^2) \in \mathbb{R}^S$.

To investigate our conjecture that more weights help better in a complex channel, we test five power levels for the bursty noise for $\sigma_b = \sqrt{P_b}\sigma$, $P_b \in \{1, {2}, {4}, {8}, {16}\}$.
From \tabref{BurstyTable}, we see that as the bursty noise power increases, the degradation of the entangled neural decoder compared to the full version also increases.

\begin{table}[!htp]\centering
\captionsetup{font=small}
\scriptsize
\begin{tabular}{cc}\toprule
\textbf{Bursty noise power} &\textbf{Degradation (dB)} \\\midrule
$\sigma^2$ &0.5 \\ 
$2\sigma^2$ &0.7 \\
$4\sigma^2$ &1 \\
$8\sigma^2$ &1.2 \\
$16\sigma^2$ &1.5 \\
\bottomrule
\end{tabular}

\caption{
As the channel worsens, with higher busty noise power, the degradation due to entangled weights (from RNN-FW to RNN-SS for NNMS decoder at BER $10^{-5}$) becomes higher.}
\label{BurstyTable}
\vspace{1mm}
\end{table}

\begin{figure}[t!]
    \centering
 	\includegraphics[width=1.0\linewidth]{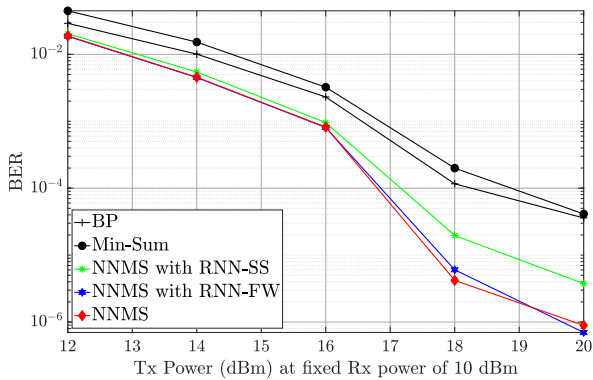}
 	\captionsetup{font=small}
 	\caption{Over The Air testing: NNMS achieves a BER of $10^{-5}$ at 1.3 dBm lower Tx power compared to RNN-SS for BCH(63,36).}
 	\label{fig:ota_tx_pow}
\end{figure}

\textbf{Over-the-Air channels.\ }Apart from the previous synthetic channels considered, we train and test the NNMS decoder in a real multi-path environment by using two USRP N200 series RF-transceivers with 1 antenna each communicating over the air. The antennas are placed at a distance of 5 meters with no direct line of sight. We generate random data, encode it using BCH(63,36) code, and modulate using BPSK. We add a synchronization preamble and transmit over the channel. At the receiver, we capture the frames, correct for frequency offset, and perform channel equalization. We then demodulate the data to estimate the LLR. Once the data is collected, the training and inference procedure is the same as other channels. \figref{fig:ota_tx_pow} shows that for a BER of $10^{-5}$, NNMS requires much lower Tx power compared to RNN-SS which has only 2 weights.

\begin{figure}[ht]
    \centering
 	\includegraphics[width=1.0\linewidth]{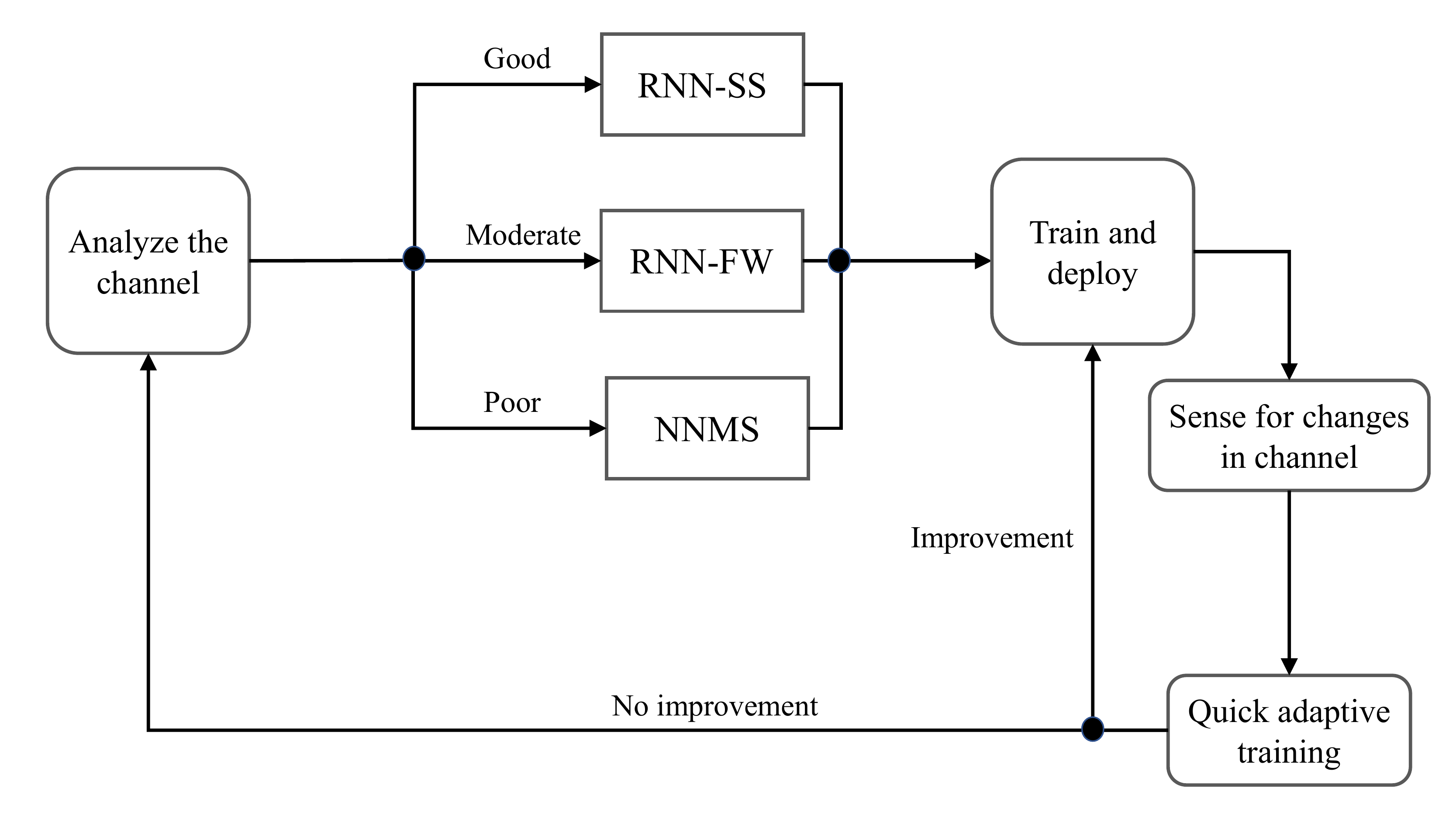}
 	\captionsetup{font=small}
 	\caption {Adaptivity of neural min-sum decoders: The model complexity can be chosen and adapted optimally on the go based on channel variations.}
  	\label{fig:adaptive_model}
  	\vspace{-3mm}
\end{figure}

These trends across different channels clearly demonstrate that the trade-off of entangling weights varies considerably with channel conditions. Thus, instead of fixing the number of weights,  we propose an adaptive framework depicted in~\figref{fig:adaptive_model}, where the model complexity is chosen based on the channel conditions. We leave quantifying the channel conditions for future work.

\section{Robustness and Adaptivity}\label{sec:robustness}

\begin{figure}[t!]
    \centering
 	\includegraphics[width=1.0\linewidth]{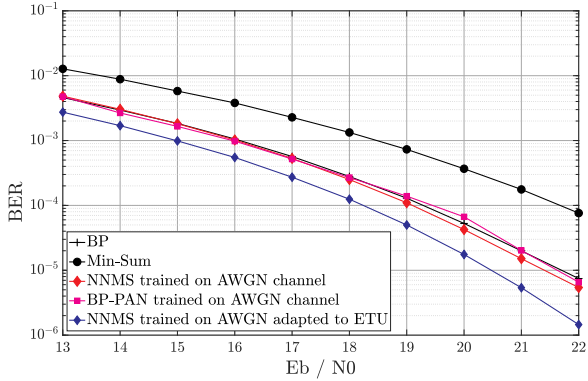}
 	\captionsetup{font=small}
 	\caption{Robustness and Adaptivity: NNMS trained on AWGN is robust and still outperforms the min-sum decoder on the ETU channel for BCH(63,36). 
    Further, with just $5\%$ of additional training on ETU data, NNMS adapts to ETU.}
  	\label{fig:nams_robust}
  	\vspace{-3mm}
\end{figure}

Since the performance of the decoders varies considerably with the channel conditions, it is important for the learned decoders to be robust. We test the robustness by training a decoder on the AWGN channel and testing on the ETU channel. \figref{fig:nams_robust} shows that when the channel changes from AWGN to ETU, the RNN-FW decoder still outperforms the original min-sum decoder. 

Even though neural min-sum decoders are robust to new channels, we still observe degradation in performance compared to the decoder trained on the true channel. To alleviate this, we propose fine-tuning the decoder to the new channel using a small amount of training data. We see from \figref{fig:nams_robust} that with just $5\%$ of additional training, the NNMS decoder adapts well to the new channel, matching the performance of the decoder trained fully on the ETU channel. This demonstrates that the neural decoders are adaptive.

This shows that while the NNMS decoder trained on the AWGN channel learns to correct the effect of short cycles, additional training on newer channels introduces the capability to correct channel effects as well.

It is interesting to note that the performance of BP-PAN\cite{lian2019} degrades noticeably when the decoder trained on AWGN is used on the ETU channel. We attribute this to the strong dependency of BP-PAN on the SNR values, which changes from AWGN to ETU for a given BER. To alleviate this, we  scale the SNR range according to the BER to match with the AWGN channel. This  improves the performance of BP-PAN trained on AWGN significantly, to match that of BP-PAN trained on ETU. However, even with this transformation, the performance is still worse than NNMS. This shows that while BP-PAN learns very well on a simple channel with very few parameters, NNMS takes advantage of the larger number of weights and generalizes well, making them more robust.

\section{Theoretical Analysis and Gaussian Approximation}\label{sec:GA}

In this section, we explore finding good normalization factors using {\em analytical} approaches and compare the performance of {analytical} and {neural} approaches. 

We propose a novel formulation using Gaussian approximation analysis, where we assume that the sum of incoming messages to the variable nodes is approximately Gaussian, as supported by empirical evidence for AWGN channels\cite{gamal2001}. Extending this assumption, we find the optimal normalization factors that minimize the probability of error for a given SNR for ETU channels. 

Specifically, we consider the  RNN-FW version of the neural min-sum decoder. We restrict the analysis to one iteration. The posterior belief at VN $v$ is thus given by
\begin{equation}
    o_{v} = l_v + \left(\sum_{c' \in N(v)} w_{c',v}*\mu_{c',v} \right) \mbox{        },
\end{equation}\label{eq:nams_final_chk}
where $\mu_{c',v}$ is belief from CN $c'$ to VN $v$ and $w_{c',v}$ denotes the  corresponding weight. Since we are dealing with only one iteration of the decoder, we can assume the incoming messages to the variable node to be independent and ignore the effect of cycles, approximating $o_v$ to the sum of independent Gaussian random variables. The resultant distribution is given by  $\mathcal{N}(\gamma_v + \sum_{c' \in N(v)} w_{c',v}\gamma_{c',v},\ \sigma^2_v +\sum_{c' \in N(v)} w^2_{c',v}\sigma^2_{c',v})$, where $\gamma$ and $\sigma^2$ denote the mean and variance of corresponding beliefs respectively. The probability of error is given by 
\begin{align*}\label{eq:perr_chk}
    P_{\textnormal{error}} = Q\left(\frac{\gamma_v + \sum_{c' \in N(v)} w_{c',v}\gamma_{c',v} }{\sqrt{\sigma^2_v + \sum_{c' \in N(v)} w^2_{c',v}\sigma^2_{c',v}}}\right).
\end{align*}

We then {\em analytically} 
find the set of optimal weights $w_{c,v}$ which minimize the probability of error, i.e., maximize the argument in the Q function. 

\begin{figure}[t!]
    \centering
 	\includegraphics[width=1\linewidth]{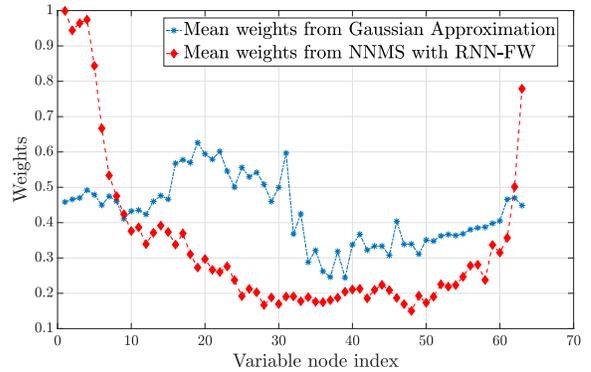}
 	\captionsetup{font=small}
 	\caption{The weights from Gaussian Approximation deviate noticeably compared to NNMS for BCH(63,36) at SNR 8 dB in ETU channel. }
 	\label{fig:gauss_weights_ent1}
\end{figure}

Now, we proceed to compare these analytical weights with the weights learned from NNMS. In \figref{fig:gauss_weights_ent1}, we plot the mean weights across the edges for each variable node, from which we see that the weights from Gaussian approximation do not capture the effect of cycles across the variable nodes and deviate from the weights learned by the NNMS decoder. 

\begin{figure}[ht]
    \centering
 	\includegraphics[width=1\linewidth]{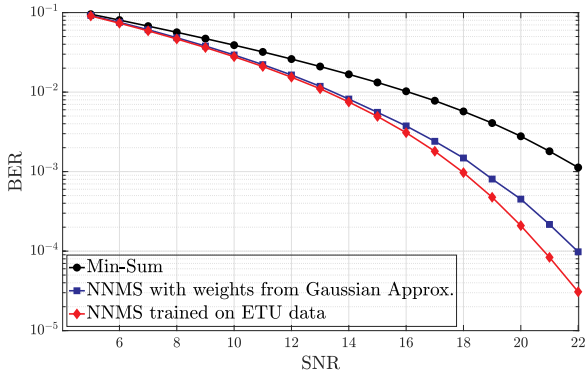}
 	\captionsetup{font=small}
 	\caption{BCH(63,36) ETU: While the Gaussian approximation approach provides reliability gains, it is still considerably worse ( > 1 dB at BER $10^{-4}$) compared to NNMS, after 1 iteration of min-sum. }
 	\label{fig:gauss_weights_ent1_ber}
\end{figure}

In \figref{fig:gauss_weights_ent1_ber} we plot the BER performance of the normalized min-sum decoder with weights obtained via Gaussian approximation and compare it against min-sum and NNMS decoders. 

We observe that while the Gaussian approximation approach provides non-negligible gains, it is still significantly poorer compared to NNMS. 
This shows that while approximating the incoming beliefs to be Gaussian is applicable for AWGN channel \cite{gamal2001}, which makes formulating the analytical solution tractable, the same cannot be extended to ETU channels. 

Alternative to analytical approaches, the problem of finding optimal weights can also be solved using backpropagation of the error for any channel conditions even for a large number of weights, demonstrating the advantage of neural decoders over the conventional approach.

\section{Conclusion and Remarks}\label{sec:conclusion}
In this work, we provide an interpretation of the neural min-sum decoders. We provide empirical evidence showing that the weights learned by the decoder are strongly influenced by the number of short cycles present in the Tanner graph. We show that the learned weights attenuate the effect of these cycles to improve the reliability of the posterior LLRs and contribute to the robustness of the decoders across channels. 

Further, we studied the complexity vs performance trade-off for these decoders across various channels. Through our simulations, we show that more weights are needed for complex channels to fully realize the gains while fewer weights are sufficient for simple channels. We show that the weights learned are robust to channel variations and can be quickly adapted to newer channels. Additionally, we demonstrate the performance of the neural min-sum decoders on practical channels using SDRs. 

 Finally, we propose a novel Gaussian approximation analysis for the neural min-sum decoders and study its performance. We show that for complicated channels, neural decoders lead to much better performance than the analytically driven weights under the Gaussian approximation. 
 
 There are several interesting open problems. The first is to understand more about quantifying the channel conditions and simplifying the selection of model complexity. Additionally, it would be interesting to establish connections between the choice of hyperparameters, the channel conditions, and the structure of the code, which could result in faster learning of the weights. 

\section*{Acknowledgment} 
This work was partly supported by ONR Award N00014-21-1-2379, ARO Award W911NF-23-1-0062, NSF Award CNS-2008824, and the 6G$@$UT center within the Wireless Networking and Communications Group (WNCG) at the University of Texas at Austin. 

\vspace{-.8em} 
 \medskip
 \small
 \bibliographystyle{IEEEtran}
 \bibliography{bibilography}

\clearpage
\normalsize

\clearpage

\end{document}